\documentclass[11pt,twoside]{article}
\usepackage{graphicx,epsfig,natbib}
\usepackage{CS18}
\begin{document}
%
%
\title{Characterizing the Parents: Exoplanets Around Cool Stars}
\author{Kaspar von Braun$^{1}$, Tabetha S. Boyajian$^{2}$, Gerard T. van Belle$^{3}$, Andrew Mann$^{4}$, Stephen R. Kane$^{5}$}
\affil{$^1$Max-Planck-Institute for Astronomy (MPIA), K\"{o}nigstuhl 17, 69117 Heidelberg, Germany; braun@mpia.de}
\affil{$^2$Yale University, P.O. Box 208101, New Haven, CT 06520-8101, USA}
\affil{$^3$Lowell Observatory, 1400 West Mars Hill Road, Flagstaff, AZ 86001, USA}
\affil{$^4$University of Texas, 2515 Speedway, Stop C1400, Austin, TX 78712-1205, USA}
\affil{$^5$San Francisco State University, 1600 Holloway Ave., San Francisco, CA 94132, USA}
\begin{abstract}
The large majority of stars in the Milky Way are late-type dwarfs, and the frequency of low-mass exoplanets in orbits around these late-type dwarfs appears to be high. In order to characterize the radiation environments and habitable zones of the cool exoplanet host stars, stellar radius and effective temperature, and thus luminosity, are required. It is in the stellar low-mass regime, however, where the predictive power of stellar models is often limited by sparse data volume with which to calibrate the methods. We show results from our CHARA survey that provides directly determined stellar parameters based on interferometric diameter measurements, trigonometric parallax, and spectral energy distribution fitting. 
%
\end{abstract}
\section{``Why?" and ``How?": Introduction and Methods}
Essentially every astrophysical parameter of any exoplanet is a function of its equivalent host star parameters (radius, surface temperature, mass, etc.). {\bf You only understand any exoplanet as well as you understand its respective parent star.} The main purpose of the presented research is to directly characterize exoplanets in orbits around their hosts and to produce empirical constraints to stellar models. 
We use infrared and optical interferometry, coupled with spectral energy distribution fitting and trigonometric parallax values, to get estimates of stellar radii and effective temperatures that are as model-independent as possible. For more details, see, e. g., \citet{boy13} and \citet{von14}. For transiting planets, using literature photometry and spectroscopy time-series data allows for the determination of model-independent planetary and stellar masses, radii, and bulk densities \citep[e. g.,][]{von12}.
\section{``So What?": Results}
\begin{figure}										
\centering
\epsfig{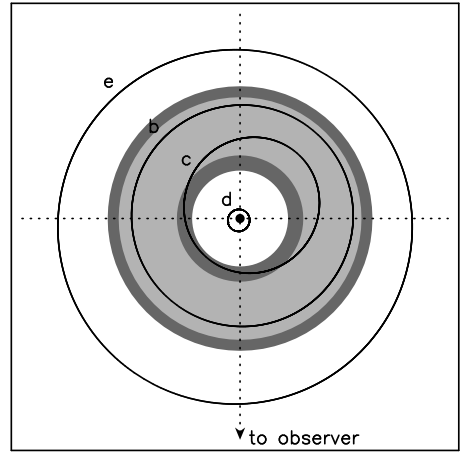} \\
\caption{Habitable zones are calculated based on our empirical values of stellar radii and effective temperatures. This plot shows the system architecture of the GJ~876 system. The HZ is shown in grey. Planets b and c spend most or all of their orbital durations in the HZ. For scale: the size of the box is 0.8 AU $\times$ 0.8 AU. Adapted from \citet{von14}.}
\label{fig:HZ}
\end{figure}

Our results provide empirically determined values for stellar radii, effective temperatures, and luminosities. They confirm the well-documented discrepancy between predicted and empirical radii and temperatures \citep[e.g.,][]{tor10, boy12,boy13} and can thus provide constraints to improvements to stellar models. They can furthermore be used to establish relations to predict stellar sizes based on observable quantities, like stellar broad-band colors, for stars too faint and/or small to be studied interferometrically \citep{boy14}.
In addition, any individual system's circumstellar habitable zone (HZ) is a function of stellar radius and effective temperature (Fig. \ref{fig:HZ}).
\section{``And what have you done for me lately?": Status}
\begin{figure*}
\plotone{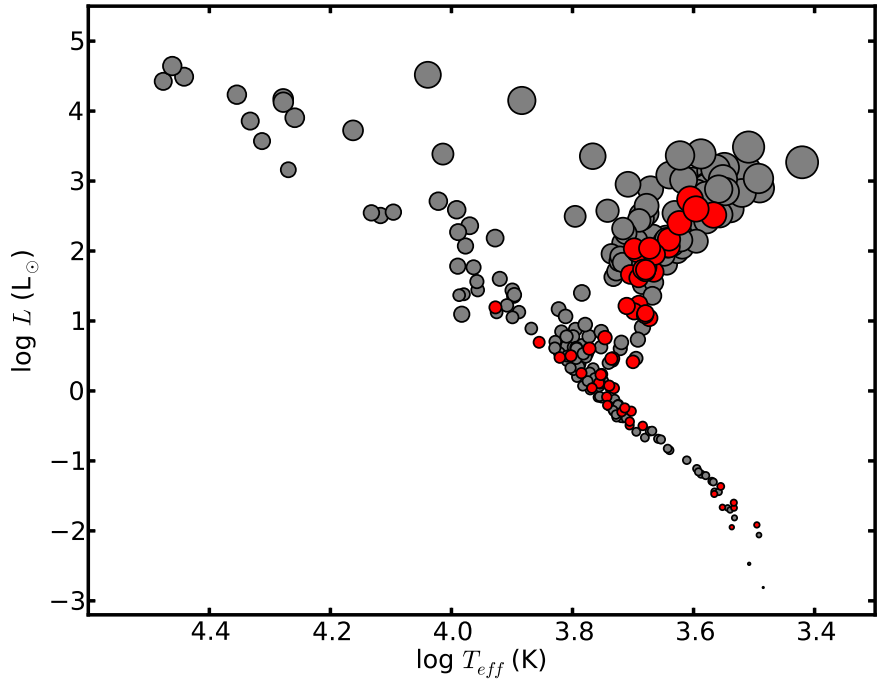}
\caption{Empirical H-R Diagram for all stars with interferometrically determined stellar radii whose random uncertainties are smaller than 5\%. The diameter of each data point is representative of the logarithm of the corresponding stellar radius. Error bars in effective temperature and luminosity are smaller than the size of the data points. Exoplanet host stars are shown in red; stars that are not currently known to host any exoplanets are shown in grey. Stellar radii data are taken from \citet{bai08, bai12, bai13, big06, boy08, boy12, boy13, dif04, dif07, hen13, ker03, lig12, ric05, van99, van09, von11a, von11b, von12, von14, whi13}. \label{fig:HRD}}
\end{figure*}

Over the course of the last 5+ years, we have been using the CHARA interferometric arrray to directly determine the stellar parameters of over 100 main-sequence stars and of around 30 exoplanet host stars, with a particular emphasis on cool stars (Fig. \ref{fig:HRD}).

\acknowledgments
{
We would like to sincerely thank the organizers for a phantastic conference. We furthermore express our gratitude to the poster judges for their thumbs up on our work. 
This research has made use of the Habitable Zone Gallery at hzgallery.org \citep{kan12}.
}

\begin{figure*}
\plotone{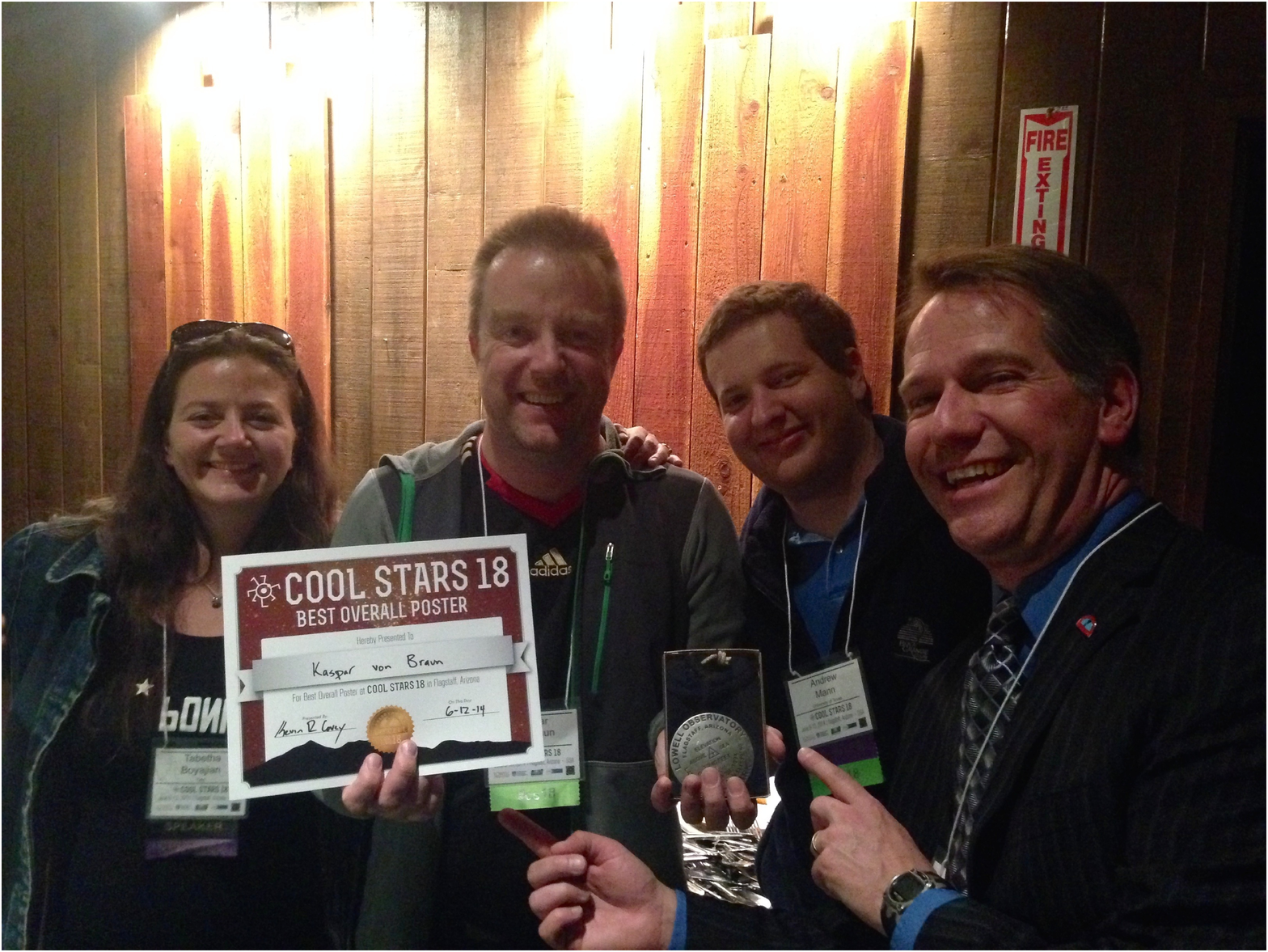}
\caption{Thank you, organizers and poster judges! \label{fig:prize}}
\end{figure*}

\end{document}